\documentclass[twocolumn,english,superscriptaddress,pre]{revtex4-1}
\usepackage[T1]{fontenc}
\usepackage[latin9]{inputenc}
\usepackage{float}
\usepackage{amsmath}
\usepackage{graphicx}
\usepackage[dvipsnames]{xcolor}

\makeatletter
\usepackage{babel}

\makeatother

\usepackage{babel}

\begin{document}
\title{Poisson's ratio and angle bending in spring networks}

\author{Nidhi Pashine}
\email{npashine@uchicago.edu}
\affiliation{Department of Physics and The James Franck and Enrico Fermi
Institutes, University of Chicago, Chicago IL, 60637.}

\author{Daniel R. Reid}
\affiliation{Pritzker School of Molecular Engineering, University of Chicago, Chicago IL, 60637.}

\author{Meng Shen}
\affiliation{Pritzker School of Molecular Engineering, University of Chicago, Chicago IL, 60637.}

\author{Juan J. de Pablo}
\affiliation{Pritzker School of Molecular Engineering, University of Chicago, Chicago IL, 60637.}

\author{Sidney R. Nagel}
\affiliation{Department of Physics and The James Franck and Enrico Fermi
Institutes, University of Chicago, Chicago IL, 60637.}

\begin{abstract}
The Poisson's ratio of a spring network system has been shown to depend not only on the geometry but also on the relative strength of angle-bending forces in comparison to the bond-compression forces in the system. Here we derive the very simple analytic result that in systems where the spring interaction strength is equal to the bond-reorientation interaction, the Poisson's ratio identically goes to zero and is independent of the network geometry.
\end{abstract}
\maketitle

Manipulating the individual bonds in elastic networks allows the design of mechanical metamaterials with unusual and unique properties~\cite{Goodrich2015,Rocks2017,Yan2017,liu2019realizing}. The simplest models for such networks consist of a single type of interaction where each bond in the network is a harmonic, central-force spring.
In two-dimensional laboratory experiments on physical networks, there is at least one other interaction that costs energy: the reorientation of the bond directions at their respective nodes. There has been some work that has examined how the the inclusion of such interactions would affect a metamaterial system's elastic properties~\cite{Reid2018, Reid2019, coulais2018multi, Rens2019, Francois2017}.   

One process of particular interest is the ability to vary the Poisson's ratio: $\nu \equiv - \epsilon_{trans} / \epsilon_{axial}$ where $\epsilon_{trans}$ is the transverse strain induced by an axial applied strain $\epsilon_{axial}$. Notably, there has been significant interest in designing materials with a negative Poisson's ratio, also known as auxetic materials~\cite{bertoldi2010negative, Reid2018,Reid2019, ren2018auxetic}.
In an isotropic network with only bond-compression forces, it was previously demonstrated that an effective strategy for tuning the Poisson's ratio to any desired value in the theoretically allowed range, $-1 <\nu<+1$, is by pruning a relatively small fraction of specially-selected bonds~\cite{Goodrich2015}. In particular, $\nu \rightarrow -1$ can be obtained by sequentially removing at each iteration the bond that contributes the greatest amount to the bulk modulus. 

However, even a small addition of an energy cost to reorient a bond's direction results in a substantial reduction in the ability to tune in desired material response. Many more bonds need to be removed to attain a desired value of $\nu$; moreover, the range of achievable $\nu$ is significantly constrained~\cite{Reid2018,Rens2019}. 

\begin{figure}
    \centering
    \includegraphics[scale=.44]{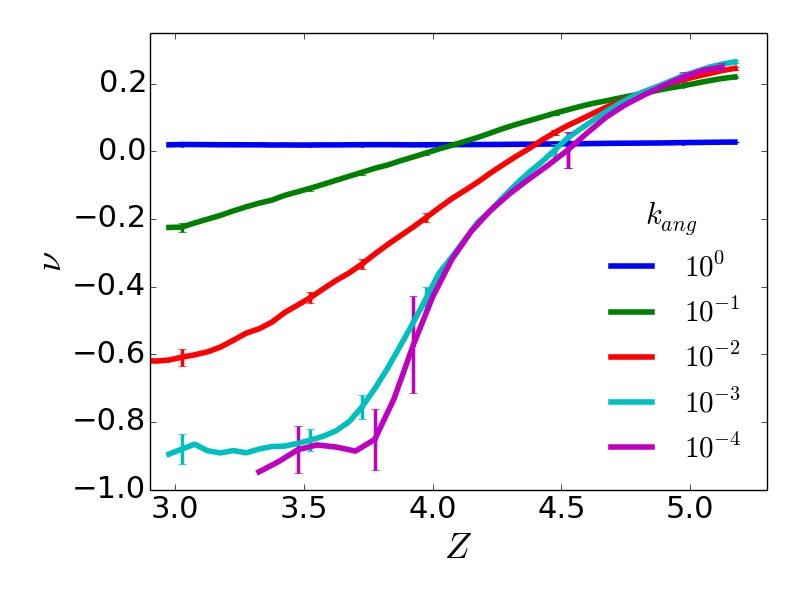}
    \caption{Poisson's ratios resulting from pruning 500-node disordered networks with $k_{ang}$ values which range from $10^{-4}$ to $10^0$. Networks have an initial $Z = 5.2$ and bonds are selectively pruned in a way that lowers the Poisson's ratio. For low values of $k_{ang}$, the Poisson's ratio of the network is highly tunable. For $k_{ang}=1$, the Poisson's ratio is $\sim 0$ and independent of $Z$.}
    \label{fig_data}
\end{figure}

This is illustrated in Fig.~\ref{fig_data}, reproduced from ~\cite{Reid2018}, where the Poisson's ratio is plotted versus the average coordination number $Z$.
In that work $Z$ was chosen to have an initial value, $Z=5.2$ and was then decreased by pruning bonds sequentially in order to minimize the Poisson's ratio. Each curve in this figure shows the results obtained with different values of the interaction parameters: $k_{L}$ (kept constant at $k_{L}=1$) which measures the restoring force to bond compression and $k_{ang}$ which measures the restoring force when the angle between two bonds emerging from the same node is varied from its equilibrium value. This plot shows how easily a disordered network can be tuned as a function of $k_{ang}$. 
For small $k_{ang}$, the Poisson's ratio can be minimized to approach $\nu = -1$ below $Z = 3.4$.  However, when $k_{ang} = 1$, $\nu$ is very close to $0$ and completely independent of $Z$. Such a network is not tunable at all. The trend shows that introducing angle-bending restoring forces makes the elastic properties of networks harder to tune.

Here we show why, by increasing the relative strength of bond-reorientation with respect to bond-compression interactions, the ability to tune the network becomes increasingly difficult. In particular, we show that, independent of the network geometry, there is one particular condition between those two interactions at which the material must have a vanishing Poisson's ratio, $\nu = 0$.

Our model for two-dimensional spring network systems includes two different interactions. Stretching or compressing a bond along its length produces a restoring force:
\begin{equation}
    F_L = - k_L dL.
\end{equation}
Rotating a bond about a node likewise produces a restoring force:
\begin{equation}
    F_{\theta} = - k_{\theta} L d\theta.
\end{equation}
(Note here that we are defining the angular restoring force with respect to the coordinate axes and \textit{not} with respect to its neighboring bonds on the same node.  This choice leads to a significant simplification.  We note that this is not the same expression as was used in~\cite{Reid2018} where the forces depend on the angle between a bond and the director at that node. Additionally, because of the way potential energy was defined, the coefficient for angle bending potential in reference~\cite{Reid2018} has different dimensions than $k_{\theta}$.)  Both of these contributions are harmonic potential interactions and their strengths are set by the spring constants $k_L$(bond compression) and $k_{\theta}$(bond orientation).  

\begin{figure}
    \centering
    \includegraphics[scale=.27]{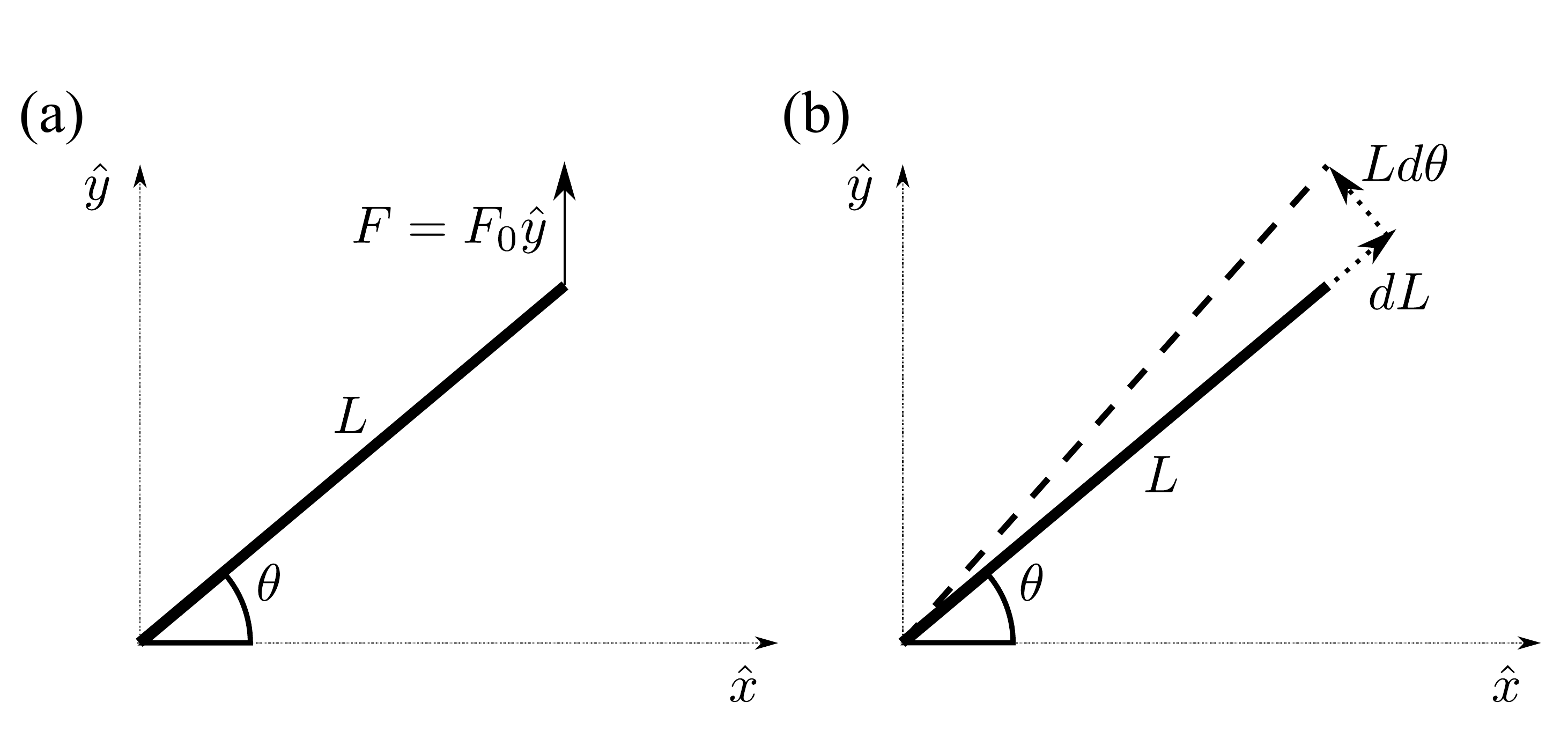}
    \caption{(a) Force applied on bond $L$ along $\hat{y}$. (b) $F$ results in change in length along the bond $(dL)$ as well as perpendicular to the bond $(Ld\theta)$}
    \label{figl}
\end{figure}




If we consider a bond of length $L$ oriented at an angle $\theta$ as shown in Fig.~\ref{figl}a, we can find the condition for which a network made of such bonds will have a zero Poisson's ratio: $\nu = 0$. In order for $\nu = 0$, a sufficient condition is that each individual bond should have no motion in the direction perpendicular to the applied force. 

Without loss of generality, we consider a force $F_0 \hat{y}$ being applied on the bond. In order to have a zero Poisson's ratio, this force should result in no change in length along the $\hat{x}$ direction. 
We can decompose $F_0$ in the directions along and perpendicular to the bond.

\begin{equation}
    F_{L} = F_0 \sin\theta
    \label{sin}
\end{equation}
\begin{equation}
    F_{\theta} = F_0 \cos\theta
    \label{cos}
\end{equation}
dividing Eq.~\ref{cos} by  Eq.~\ref{sin} gives,
\begin{equation}
    \tan\theta = \frac{F_L}{F_{\theta}} = 
    \frac{k_L}{k_{\theta}}\frac{dL}{L d\theta}
     \label{ratio1}
\end{equation}
For the Poisson's ratio to be zero, we need that there be no net motion along $\hat{x}$.  The geometry depicted in Fig.~\ref{figl}b shows that this leads to:
\begin{equation}
dL \cos\theta - L d\theta \sin\theta = 0.
\end{equation}
Combining this with Eq.~\ref{ratio1} leads to:
\begin{equation}
\frac{dL}{L d\theta} =  \tan\theta =  \frac{k_L}{k_{\theta}}\frac{dL}{L d\theta}
\end{equation}
This implies that for $\nu = 0$, it is sufficient to have
\begin{equation}
    \frac{k_{L}}{k_{\theta}} = 1. 
\end{equation}
This is true irrespective of the structure of the network. 

This is an extremely simple and general result that holds for any geometry of two-dimensional networks.  For any given spring network, if we can modify the relative strengths of angle-bending and bond-compression interactions, we can always achieve the result that the Poisson's ratio of the network is exactly zero. 

The simplicity of the expression stems from our use of $k_{\theta}$ as the spring constant for changes of the direction of a bond with respect to the laboratory-frame axis, $\hat{x}$.  This of course is slightly different from the restoring force computed due to the change in angle between two adjacent bonds emerging from a node. In reference~\cite{Reid2018}, angular changes at each node were measured with respect to a director that can itself rotate in a way that minimizes the angular potential energy. Additionally, in reference~\cite{Reid2018}, the value of $k_L$ for each bond is slightly different depending on the length of the bond.
However, as shown in Fig.~\ref{fig_data}, the results do not differ appreciably if each node has many bonds emerging from it. In addition, 
the small difference between the data shown in Fig.~\ref{fig_data}, where the Poisson's ratio is \textit{very close} to $0$ and our calculation where the Poisson's ratio is \textit{exactly} $0$ for $\frac{k_{L}}{k_{\theta}} = 1$ also comes from how angle change is taken into account in these two cases. 

This result explains why it is more difficult to tune the Poisson's ratio of a network when bond-reorientation forces are present in comparison to than when they are absent. Indeed, our analysis above shows that if $\frac{k_{L}}{k_{\theta}} = 1$, it is impossible to change the Poisson's ratio at all.  The closer the spring-constant ratio is to this value, the harder it is to get a variation in $\nu$.

This work was supported by the National Science Foundation (MRSEC program NSF-DMR 2011854) and the Center for Hierarchical Materials Design (Grant No. 70NANB19H005).


\end{document}